# A race model for singular olfactory receptor expression

Brian E. Kolterman, Ivan Iossifov, and Alexei A. Koulakov

*Cold Spring Harbor Laboratory, Cold Spring Harbor, NY 11724*

**ABSTRACT**

In vertebrates, olfactory sensory neurons choose only one olfactory receptor to produce out of ~2000 possibilities. The mechanism for how this singular receptor expression occurs is unknown. Here we propose a mechanism that can stochastically select a single gene out of a large number of possibilities. In this model, receptor genes compete for a limited pool of transcription factors (TFs). The gene that recruits a target number of TFs is selected for expression. To support this mechanism, we have attempted to detect repeated motifs within known sequences of mouse olfactory receptor promoters. We find motifs that are significantly overrepresented in olfactory versus other gene promoters. We identify possible TFs that can target these motifs. Our model suggests that a small number of TFs can control the selection of a single gene out of ~2000 possibilities.

In the mammalian olfactory system, odorants are detected by olfactory sensory neurons (OSNs). Each of these neurons chooses only one olfactory receptor (OR) protein to express out of a large repertoire of possible genes, a phenomenon called singular olfactory receptor expression [1, 2]. Because there are about 1000 OR genes present in the haploid mammalian genome, combined with the fact that OR genes are expressed in a monoallelic fashion, there are ~2000 possibilities to choose from for every cell [3]. The mechanism of singular gene selection is not understood.

It is proposed that when a functional OR gene is produced, it inhibits the expression of other OR genes through a negative feedback mechanism [4]. It is thought that due to this mechanism the co-expression of two functional genes is not observed [4-6]. On the other hand, when a non-functional OR gene is inserted into the mouse genome, the neurons selecting it can switch to a functional gene [4, 5]. This implies that two OR genes can be transcribed at the same time when only one of them directs the production of a functional receptor protein. Within the olfactory genome, there are a number of OR pseudogenes which have accumulated enough mutations such that they no longer produce a functional receptor protein. OR pseudogenes can be transcribed as seen from mRNA detected in OSNs [7], and be co-expressed with functional genes. It was thus suggested that the feedback mechanism is engaged which stops the selection process only once a functional OR gene is expressed [4].

OR genes are found in around 100 clusters containing 1 to ~100 genes each and distributed throughout the genome. Some clusters of OR genes are bound in a chromatin configuration that inhibits expression [8]. Occasionally, one (or more) of these clusters may be activated, in which case there may be ~100 choices of OR genes to make instead of ~2000. Although the choices for gene selection are biased by clustering, chromatin remodeling, and zonal organization within the olfactory epithelium [8-10], the problem of singular OR choice is not resolved by these mechanisms.

OR choice could be determined by a genetic rearrangement, as in the case of antigen receptor expression in developing B cells in the immune system [11]. This possibility however has been ruled out experimentally [12]. It is likely therefore that a single gene is selected by a transcriptional network within each cell. Promising regulatory elements for this network are the *cis* acting H- or P- elements [13, 14]. These appear to have an effect only on nearby OR genes [15]. Because analogs of H- and P- elements are not known for most of the OR genes, the plausible candidates lie within the promoter region of each OR gene itself.

Here we study computationally a possible mechanism of singular OR gene choice by a transcriptional network. In one possible model, the expression of each OR gene is controlled by a distinct transcription factor (TF). This would mean that a large portion of the genome is devoted to OR specific TFs, something that is not plausible. This leaves a mechanism based on a small number of TFs which governs the expression of the majority of OR genes. In this paper, we demonstrate a possible mechanism of stochastic OR gene choice which exhibits singular expression based on the binding of TFs to OR promoter regions. In this model, the promoter regions compete for TFs from a common but limited pool. The promoter that reaches a certain number of bound TFs activates the expression of the corresponding OR gene. Therefore we call this mechanism the "race" model. Motivated by this model, by analyzing a large ensemble of OR promoter sequences, we find a TF binding motif that is enriched specifically in mouse OR promoters.

**RESULTS**

**Race model for singular OR gene expression.** In our model, each of the OR genes competes for a limited pool of TFs (Figure 1). We assume that each OR promoter region can bind several TF molecules. A gene in which the number of TFs bound to its promoter equals or exceeds some predetermined number is chosen to be transcribed (Methods). If the number of TFs bound drops below this threshold, the gene is no longer transcribed. Once an OR is produced, if it is a functional protein, a feedback mechanism is initiated which inhibits the rate of TF production, reducing it to a smaller, but non-zero basal level compared to the rate during the choice phase of the model. As the concentration of unbound TFs drops, it will become increasingly more difficult for other genes to bind enough TFs in order to be expressed. The first OR selected can maintain stable expression, however, despite the decrease in the level of TF.

The stability of OR choice is ensured in our model by the cooperative binding of TFs to the promoter region (Figure 2). This means that the rates of TF association and dissociation from the promoter are different and depend on the number of already bound TFs. For simplicity, we assume that the rate of TF association does not depend on the number of TFs already bound to a given promoter. The rate of TF dissociation is lower for promoters with higher TF occupancy. This effect reflects the affinity between TFs that are bound to DNA in close proximity to one another. This type of binding has been studied in other systems such as the *lambda* repressor [16]. The overall effect of cooperative binding is that the rate of TF accumulation is larger for promoters with larger TF occupancy (Figure 2A). Cooperative binding implements a winner-takes-all type of competition in the system containing many promoters, leading to the selection of a single active gene when the threshold for transcription is reached. At the same time, the unselected ORs do not have an opportunity to be expressed, because for such ORs, at the basal level of TF production mediated by the negative feedback mechanism, the rate of TF dissociation is higher than the rate of association (Figure 2B).

Our simulations show that this model can select a single OR out of a large ensemble of available genes (Figure 3A). The selection occurs stochastically depending on the random conditions within each cell. We find that singular OR expression in our model is also stable. This means that a single OR gene choice is unlikely to be changed, with rare exceptions (Figure 3A). The reason for this stability is two-fold. First, the negative feedback mechanism reduces the level of TF once the functional OR gene is selected, preventing other genes from being expressed (Figure 2B). Second, despite the low levels of available TF, the active promoter remains active, due to the stability of TF binding introduced by the cooperativity (Figure 2B). This observation motivates the proposal that several TFs per promoter are needed to activate an OR gene. These results can be replicated if the number of genes is increased to 2000 and for other values of parameters. The particular values used in Figure 3 are chosen for easier display.

We also model OR choice with the condition of non-cooperative TF binding (Figure 3B). In the non-cooperative binding case, there are moments in time when the number of TFs bound does exceed the threshold required for expression (in this case 4). However, this choice is never maintained and therefore lacks stability. On the other hand, the cooperative binding case (Figure 3A) demonstrates very stable maintenance of the required number of factors above the threshold over long periods of simulated time.

The stability seen in OR choice also extends to the case when pseudogenes are included (Figure 4A-D). For this simulation, cooperative binding of TFs is used and the OR population is split into 80 functional genes and 20 pseudogenes. The pseudogenes can be selected in the same way as the functional genes; however, unlike a functional gene, this choice does not engage the negative feedback mechanism and hence does not stabilize the choice. This does allow for various levels of pseudogene expression, but the singular choice probability of functional OR genes is unchanged (Figure 4E).

Even though TFs appear as a single type in the simulation, the model is still consistent with the possibility that there are in fact a number of different types of TFs needed to elicit a gene choice. If the TFs involved in OR gene choice that bind to OR promoter regions are heterogeneous, this model predicts their expression levels to be co-regulated and that they should exhibit cooperative binding. This possibility will be discussed later. In the simplest case, however, the requirement of multiple TFs per promoter may be satisfied if several similar binding elements are present within each promoter. This observation motivated us to search for repeated motifs in the promoter regions of OR genes sharing this type of regulatory logic. Below we explore this requirement in the mouse genome.

**Repeated promoter motifs.** Motivated by the race model, we search for repeated motifs within a known ensemble of mouse OR promoters. We studied the promoters of 1085 OR genes as identified in Clowney, Magklara [17]. We looked for motifs of length 7 to 11 with varying numbers of mismatches and found the most significant non-trivial result to be a motif of length 7 with 1 mismatch (see Methods for details of analysis).

The consensus sequence generated by the motifs in all of the 1085 OR promoters is the palindrome: AAATAAA (Figure 5). On average, this motif is repeated 12 times in each OR promoter. To evaluate the significance of this result, we applied the same algorithm that was used to find repeated motifs in OR promoters to 3 different collections of promoters including randomly generated sequences and randomly selected mouse gene promoters. In each case we looked for the maximally repeated motif; not necessarily the same as the OR promoter motif (Figure 6). We find that the number of repeats seen in OR promoters is significantly higher than the number of repeats in randomly chosen mouse promoters (p< $10^{-197}$ ). Sequences generated by randomly shuffling the OR promoter nucleotides, which retains their GC fraction, yield even fewer repeats (Figure 6). Our finding of an enriched repeated motif in OR promoters supports the assumptions of the race model.

**TF candidates.** Having a candidate motif, we performed a search of the Jaspar transcription factor database [18] for possible TFs that can bind to it. To perform the search we downloaded the entire TF database and calculated the correlation between our discovered motif and the position weight matrices in the database. Table 1 lists the top 10 TFs with the highest correlation. Of these candidates there are several of particular interest.

Some of these molecules display behaviors consistent with their involvement in OR expression. The TF with the highest correlation is Foxj1. This is an intriguing possibility as it is expressed in the olfactory epithelium although not in olfactory sensory neurons. It is expressed in non-ciliated cells in the olfactory epithelium as well as in motile ciliated cells throughout the respiratory tract [19]. Knockout mice of Foxj1 show formation of olfactory receptor neurons and axonal projection into the olfactory bulb but with no glomerular formation [20] consistent with the hypothesis that Foxj1 leads to a failure of olfactory receptor neurons to express OR genes. Generally it is believed that ORs are required for axonal guidance and therefore glomerular formation [21]. Another TF of interest is Foxa2, which is also expressed in the olfactory epithelium at low levels in the adult mouse [22].

**DISCUSSION**

In this paper we have proposed a mechanism that is capable of reproducing singular stochastic OR gene expression. In our model, the promoter regions of different OR genes compete for the limited pool of TFs.

The promoter that binds a certain number of TFs activates an OR gene for expression. Because our model is based on the competition for TFs by multiple OR gene promoters, we call it the "race" model.

Our model relies on three assumptions. First, we assume multiplicity of regulatory sites in each OR gene promoter, which implies that several TFs are needed to activate transcription of one OR gene. In our model this number is four, but the exact value is not clear. Second, we assume cooperativity between these TFs. We have determined that cooperative binding between the TFs in our model leads to stable OR gene expression, as opposed to non-cooperative binding which does not exhibit stable expression. Third, we assume a negative feedback regulatory mechanism that is activated by the expressed functional OR [4]. Specifically, we assume that the expression of a functional gene reduces the level of TF in the cell, preventing other ORs from being expressed.

Although the identity of the TFs involved is not clear, our model suggests several alternatives. Our model is based on multiple TFs binding to the same promoter in order to activate an OR gene. This can occur by several mechanisms. In the simplest case, several identical TF binding motifs may be present within each promoter that recruit the same TF. In an effort to explore this option, we carried out an analysis on the known set of 1085 mouse OR promoters to find the maximally repeated motif common to all. We have shown that the promoter regions of OR genes contain a repeated motif that is significantly over-represented compared to the promoter regions of randomly selected mouse genes. The most overrepresented motif (AAATAAA) was found to be repeated on average about 12 times per the 1kbp stretch upstream from known mouse OR transcription start sites. We used this motif to identify candidate TFs from a database of known binding motifs. We have found possible candidate TFs including Foxj1 and Foxa2 among others. Interestingly, these TFs are expressed by cells in the olfactory epithelium in a manner that is consistent with their potential role in OR selection. In addition, Foxj1 knockouts do not display normally segregated olfactory glomeruli [20]. The presence of repeated TF binding motifs in the OR gene promoters supports our computational model that is based on the race between different promoters in their binding and recruitment of several TFs that can be of the same type, i.e. homogeneous.

Another possibility is that several heterogeneous TFs of different type are required to bind to the same promoter to activate an OR gene. Because the number of TF types in this case is much smaller than the number of OR genes, this mechanism is consistent with the race model as long as these TFs exhibit cooperative binding. Mathematically the heterogeneous and homogeneous models are similar. There are a number of candidates that support the heterogeneous TF option. Olf-1/EBF-like (O/E) binding sites are known to exist in many OR promoters and O/E genes are expressed in the olfactory epithelium [17, 23, 24]. Knockouts of O/E-2 and O/E-3 do not form normal glomeruli but retain OSNs, suggesting a possible role in OR expression [25]. In addition to the O/E sites, there are also homeodomain (HD) and homeobox classes of motifs found in OR promoters. Two TFs belonging to these classes are known to be involved in OR expression (Lhx2 and Emx2) [26, 27]. HD and O/E sites are present in most OR promoters as well as in the H and P elements [17, 28]. Transgenic experiments in which multiple repeats of HD sites are placed upstream of an OR gene show greatly enhanced expression, suggesting that these HD motifs are important in modulating the probability of OR gene choice [28]. This result is expected in the race model because a larger number of bound TFs can confer greater stability to a given OR choice. For some mouse ORs, such as M71, as little as 161bp upstream from the transcription start site is sufficient to support choice [29]. Interestingly, this segment of M71 promoter contains the motif found in this study (AAATAAA) as well as HD and O/E binding motifs. Another interesting set of candidates supporting the heterogeneous model are the POU domain classes of TFs whose binding motifs are found to be overrepresented in OR promoters compared to other mouse gene promoters [17]. These motifs are targeted by a number of different TFs in the Brn POU domain factors (BRNF) family. It is already known that the TFs in this family can bind cooperatively [30, 31]. Finally, TFs binding to the motif found to be enriched in this study could cooperate with these other TFs in facilitating OR expression.

**METHODS**

**OR Expression Simulation**

The OR gene promoter-TF system is modeled with the following stoichiometric equations:

$$P_g^n + TF \underset{k_{-}(n)}{\overset{k_{+}}{\rightleftharpoons}} P_g^{n+1}, \quad 0 \leq n < n_{max}$$

$$k_{-}(n) = k_0 e^{-\alpha(n-1)}, \quad k_{+} = \text{const}$$

$$\emptyset \underset{k_{deg}}{\overset{k_{prod}(N_{exp})}{\rightleftharpoons}} TF, \quad k_{prod}(N_{exp}) = \frac{r_0}{\mu^{N_{exp}}},$$

$$k_{deg} = \text{const}, \quad N_{exp} = \sum_g \sum_{n=n_e}^{n_{max}} \left[ P_g^n \right]$$

Here, $k_{prod}(N_{exp})$, $k_{deg}$ are the rates of TF production and degradation respectively. $TF$ is therefore a free transcription factor available in the system for binding. $N_{exp}$ is a count of the number of functional OR genes expressed. In this model, OR expression occurs whenever the number of TFs bound to a given promoter equals or exceeds some threshold number ($n_e$). $r_0$ is the maximum rate of TF production before OR expression. $\mu^{N_{exp}}$ is the suppression factor of TF production after expression occurs. $k_{+}$ is the rate of TF association to promoter regions (taken as constant). $k_{-}(n)$ gives the dissociation rate as a function of the number of TFs bound $n$, with $k_0$ being the maximum dissociation rate. $\alpha$ controls the amount of cooperative binding between TFs. $P_g^n$ identifies whether or not the $g^{th}$ OR promoter has $n$ TFs bound up to a maximum of $n_{max}$.

To capture the stochastic nature of OR selection we simulate the discrete single molecule events. This simulation could be carried out using the finite differences method with the small time interval $\Delta t$. In this case the probability of a discrete event within the time interval is given by $k\Delta t$, where $k$ is a reaction rate. To make the simulation more efficient, we chose to simulate the system of coupled chemical equations using the Gillespie exact method [32]. The basic steps of this method are: first, we randomly choose a reaction $i$ with probability $p_i \propto k_i$, the rate of this reaction. Then, we randomly generate the time elapsed since the previous reaction by sampling from an exponential distribution with the average $t_i = 1/\sum_j k_j$. Finally, we update the system by carrying out the chosen reaction. These steps are then repeated until all reactants are extinguished or a specified time has elapsed.

For the simulation in Figure 2 we used: 100 total genes, $N_{max} = 10$, $n_e = 4$, $r_0 = 10/\text{hr}$, $\mu = 15$, $k_{+} = 5/\text{hr}$ $k_0 = 30/\text{hr}$, $k_{deg} = 10/\text{hr}$ and $\alpha = 0\ \&\ 0.6$ for non-cooperative and cooperative binding respectively. For the pseudogene simulation (Figure 3) we used: 80 functional genes, 20 pseudogenes, $N_{max} = 10$, $n_e = 4$, $r_0 = 10/\text{hr}$, $\mu = 200$, $k_{+} = 5/\text{hr}$ $k_0 = 40/\text{hr}$, $k_{deg} = 10/\text{hr}$ and $\alpha = 1.0$. The parameters were chosen to yield stable expression of a single functional gene by searching through the parameter space. The time scale for the gene choice was assumed to be of the order of hours which determined the overall scales of reaction rates. With this time scale, one unit of time in Figures 2 and 3 is approximately equal to 1/3 of an hour but varies slightly due to the randomness present in the generation of the elapsed time as mentioned above.

**Over-represented Motif Search**

The search for motifs was conducted using mouse OR genes identified in Clowney, Magklara [17]. They provide transcription start sites (TSS) for 1085 mouse OR genes in the mm9 build of the mouse genome [33]. We took 1001 base pairs upstream of the TSS as being the promoter for a given OR gene. These promoters were then split into words of size 7 to 11. All of promoters were then searched for each word allowing some set number of mismatches from 0 to 2. The reverse compliment of the words was also counted while overlapping motifs were excluded. Since we are only interested in significantly over-represented motifs in OR gene promoters, only the motif with the largest number of total repeats was given as the result.

To find the significance of this result we repeated the analysis using generated promoters. We used three types of promoter collections, each type having the same number of promoters containing the same number of nucleotides as in the OR promoter analysis. The first case contains promoters generated assuming equal probability of each nucleotide. The second case uses promoters generated randomly with the same nucleotide distribution as the native mouse OR promoters. The third and final case uses 1085 randomly selected promoters from the full complement of 48970 mouse genes as annotated in the UCSC mouse genome database, build mm9 [34]. In each case, the analysis was repeated 100 times in order to get a distribution of the maximally repeated element. In this way we were able to determine the expected mean and standard deviation for the most repeated motif assuming a normal distribution. Note, the identity of the motif in these random distributions is not important as we are only interested in the largest number of repeated elements.

# FIGURES

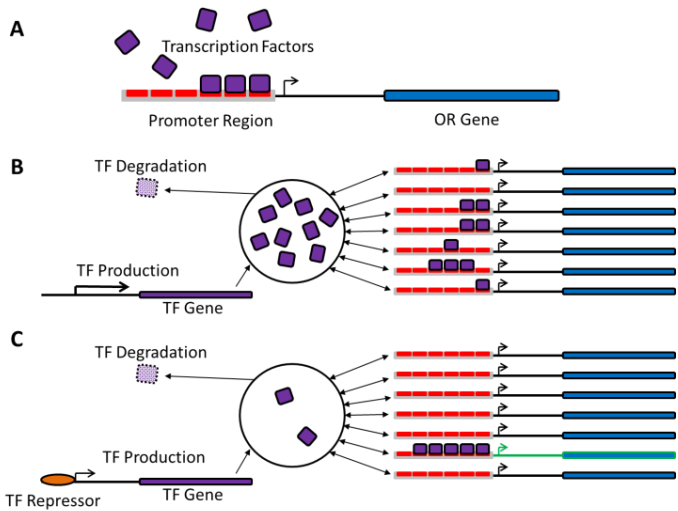

**Figure 1.** Schematic of OR TF race model. (**A**) Each OR gene has multiple TF binding motifs in its promoter region. (**B**) Each promoter competes in a race to bind some threshold number of TF from a limited pool. (**C**) The gene whose promoter wins the race is transcribed. If the gene translates to a functional OR protein, the rate of TF production is repressed and the rest of the genes are unable to bind enough TFs to be expressed.

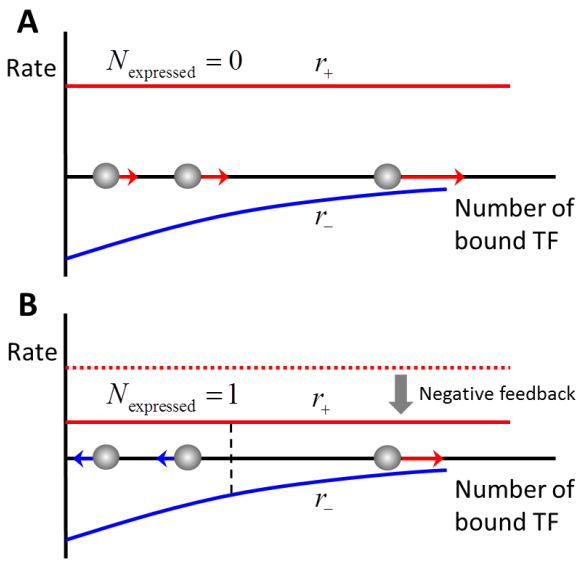

**Figure 2.** Stability analysis for TF binding. The rates of TF association/dissociation are shown as functions of the number of TFs currently bound to a promoter. Several promoters are shown in each panel for illustration. (**A**) Before an OR is expressed, all promoters (gray balls) recruit TFs, as indicated by the red arrows. The rate of TF association ($r_+$, red line) is bigger than the rate of dissociation ($r_-$, blue line). Because of cooperativity, the promoter with the largest number of TFs bound (extreme right), recruits new TFs faster. (**B**) When an OR is expressed, the rate of association ($r_+$) is reduced by the negative feedback mechanism. As a result, promoters with a small number of bound TFs, below the unstable equilibrium point (dashed line), lose their TFs. The promoter with a large number of TFs bound remains activated. With a proper choice of parameters, such a promoter is singular in most cases.

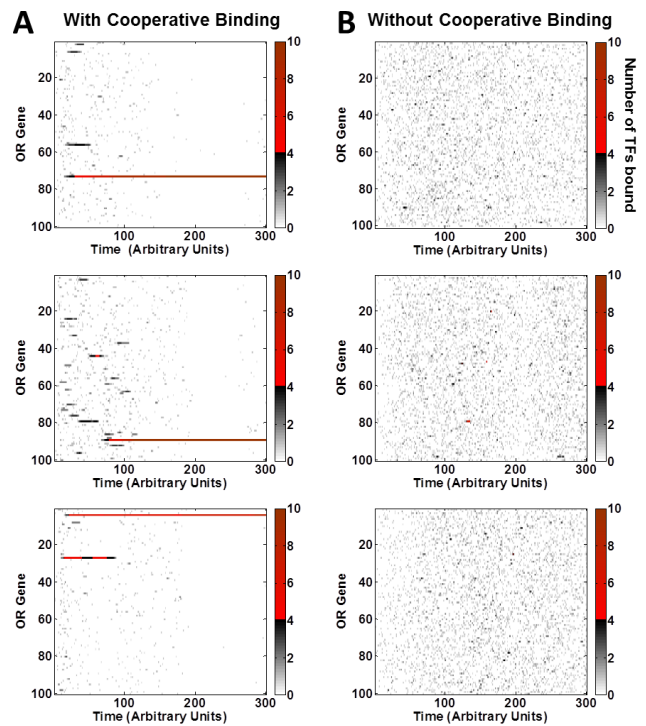

**Figure 3.** Number of TFs bound to each promoter as a function of time. Each panel shows a simulation for one neuron. OR expression (indicated in red) occurs when at least 4 TFs are bound to one promoter. (**A**) Cooperative binding of TFs leads to singular expression with stability over long time periods. (**B**) Non-cooperative binding exhibits no stable expression.

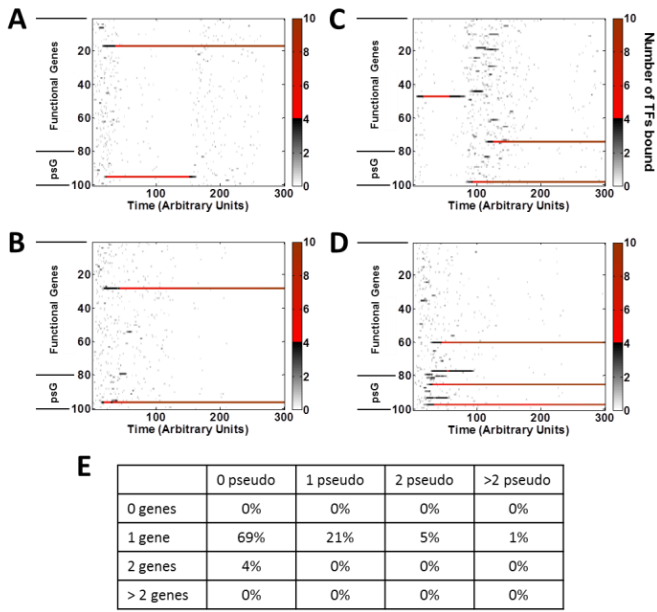

**Figure 4.** Number of TFs bound to each promoter as a function of time for 80 functional genes and 20 pseudogenes (psG) with cooperative binding of TFs. Each panel shows a simulation for a single neuron. Pseudogenes do not invoke a negative feedback mechanism to stabilize gene choice while functional genes do. OR expression occurs when at least 4 TFs are bound. Singular expression occurs for functional genes in almost all cases with occasional co-expression of pseudogenes (**B-D**). (**E**) Table of probabilities for various psG and functional gene expression level combinations determined from 1000 simulated cells.

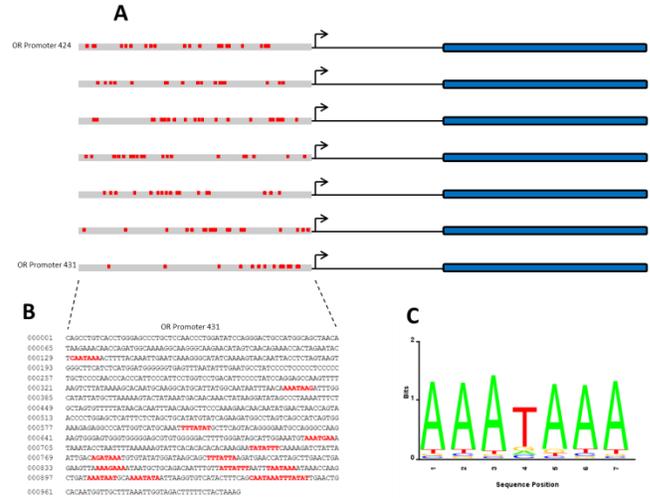

**Figure 5.** OR promoter motif. **A.** Location of the most repeated motif in a sample of 7 OR promoters in the mouse genome (build mm9). Reverse compliments are included; overlaps are excluded. **B.** Nucleotide view for one of the OR promoters. **C.** Sequence logo for the most significantly over-represented motif in all of the 1085 OR promoters.

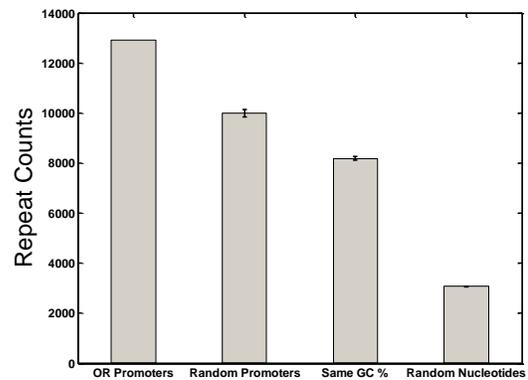

**Figure 6.** Number of maximally repeated motifs in the set of 1085 OR gene promoters and three randomly generated sets of promoters. The number of repeats in OR promoters is significantly larger than in the three random sets ($p < 10^{-197}$).

| Transcription Factor Name | Correlation to repeated motif |
|---|---|
| Foxj1 | 0.939 |
| Tcf3 | 0.887 |
| Foxa2 | 0.869 |
| Foxl1 | 0.831 |
| Hoxb13 | 0.827 |
| Hoxd13 | 0.826 |
| Hoxd10 | 0.825 |
| Foxj3 | 0.824 |
| Hoxa10 | 0.821 |
| LM193 | 0.812 |

**Table 1.** Top 10 transcription factor candidates with correlation to the over-represented motif found in OR promoters. Candidate transcription factor motifs are taken from the Jaspar database.